\documentclass[doublecol]{epl2} 
\usepackage[latin1]{inputenc}
\usepackage{textcomp}

\makeatletter
\usepackage{graphicx}
\usepackage{bm}
\usepackage{amsmath}
\usepackage{amssymb}
\usepackage[amssymb]{SIunits}

\newcommand{\Etat}[1]{{| {#1} \rangle}}

\newcommand{\mFv}[1]{\ensuremath{\Etat{ m_F }}\xspace}

\newcommand{\ve}[1]{\overrightarrow{#1}}

\newcommand{\om}{\omega}

\newcommand{\lL}{\lambda_L}

\newcommand{\be}{\begin{equation}}
\newcommand{\ee}{\end{equation}}

\newcommand{\Gif}{\Gamma_{if}}

\title{Effect of vortices on the spin-flip lifetime of atoms in superconducting atom-chips}
\shorttitle{Vortices and spin-flip lifetime in superconducting atom-chips} 

\author{G. Nogues\inst{1}\thanks{E-mail: \email{gilles.nogues@lkb.ens.fr}} \and C. Roux\inst{1} \and T. Nirrengarten\inst{1} \and A. Lupa{\c s}cu\inst{1} \and A. Emmert\inst{1} \and M. Brune\inst{1} \and J.-M. Raimond\inst{1} \and S. Haroche\inst{1,2} \and B. Pla{\c c}ais\inst{3} \and J.-J. Greffet\inst{4}}
\shortauthor{G. Nogues \etal}

\institute{                    
  \inst{1} Laboratoire Kastler Brossel, ENS, UPMC, CNRS - 24 rue Lhomond, 75005 Paris, France, EU\\
  \inst{2} Coll\`{e}ge de France - 11 place Marcelin Berthelot, 75005 Paris,
France, EU\\
  \inst{3} Laboratoire Pierre Aigrain, ENS, UPMC, CNRS - 24 rue Lhomond, F-75231 Paris Cedex 05, France, EU\\
  \inst{4} Laboratoire EM2C, Ecole Centrale Paris, CNRS, Grande Voie des Vignes, 92295 Ch{\^ a}tenay-Malabry, France, EU
}


\abstract{We study theoretically the lifetime of magnetically trapped atoms in the close vicinity of a type-II superconducting surface, in the context of superconducting atom-chips. We account for the magnetic noise created at the cloud position by the vortices present in the superconductor and give specific results for our experiment which uses a niobium film. Our main result is that atom losses are dominated by the presence of vortices. They remain however dramatically smaller than in equivalent room-temperature atom-chip setups using normal metals. 
}

\begin{document}

\maketitle

Atom chips allow to trap ultracold atomic gases in the vicinity of micron-sized current
carrying wires~\cite{TR_ZIMMERMANRMPCHIP07} or permanent magnetic
structures~\cite{TR_HINDSREVIEWFERROMAG99}. Microfabrication techniques allow to design complex trapping potentials and to realize versatile manipulation of atoms thanks to the control of currents or radiofrequency fields in the vicinity of the trapped cloud~\cite{TR_HANSCHCONVEYOR01,TR_PRENTISSBEAMSPLITTERCHIP05}. Atoms chips are now considered as a powerful toolbox that can be used for fundamental studies~\cite{TR_SCHMIEDMAYERDBLWELL05,TR_ZIMMERMANBLOCH05}, atomic interferometry~\cite{TR_REICHELCOHERENCECHIPCLOCK04,TR_KETTERLECHIPSQUEEZ07} or quantum gate implementation~\cite{TR_CALARCO_CHIPGATE_00}.

\begin{figure}
 \centering
\includegraphics[width=6cm]{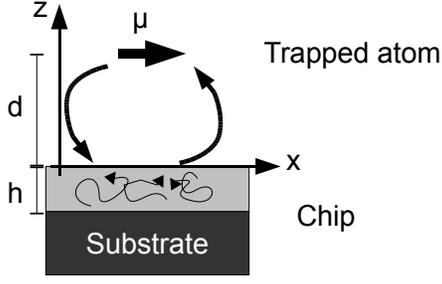}
  \caption{An atom in an initial state $|i\rangle$ is trapped at position $(0,0,d)$ in vacuum near a film of metal or superconductor of thickness $h$. We assume that the substrate under the film is dielectric and does not affect the atomic lifetime. At the level of the atomic cloud, an external magnetic field sets the quantization axis along the $x$ direction.
}
 \label{fig:geom}
\end{figure}

In many such experiments, atoms are required to be very close to the surface of the trapping structures. Unfortunately, additional losses from the trap are experimentally observed in these conditions~\cite{TR_HINDSEXPTRAPLOSS_03,TR_VULETICCHIPNOISE04}. Johnson-Nyquist current noise in the trapping metallic wires produce magnetic field fluctuations at the position of the atoms, which can induce Zeeman transitions towards untrapped magnetic sublevels. This phenomenon is strongly enhanced in the near-field of conductors for typical spin-flip  radiofrequencies (in the MHz range)~\cite{TR_HINDSREKDAL_FLIPCALCULATION_04,TR_HENKELMETALNOISE05}. The typical geometry of these experiments is presented in Fig.~\ref{fig:geom}. A possibility to overcome these difficulties consist in using cryogenic atom-chips made of superconducting materials, for which dissipation at RF frequencies, and hence fluctuations, are dramatically reduced~\cite{TR_HINDS_LIFETIMECONSUPERCON_05}. Successfull operation of superconducting atom-chips has been reported~\cite{ENS_CHIPSUPRA06,TR_SHIMIZUPERMANENTCHIP07,ENS_BECSUPRA08}, with the aim of developping new hybrid atomic--solid-state systems. Concurrently, theoretical studies have made quantitative prediction for the lifetime increase with respect to normal metals. However they strongly depend on the model of superconductivity which is used in the calculation. Most recent  articles agree to predict an enhancement of at least 6 orders of magnitude~\cite{TR_REKDALSUPERCONDNOISE06, TR_HINDSELIASHBERG07, TR_REKDAL_ELIASHBERG_07}.

In this Letter, we emphasize the importance of the vortex dynamics in the superconducting material on the atomic losses. In current atom-chip experiments, DC magnetic fields of the order of 10-100~G are applied orthogonally to thin layers of type-II superconductors. Hence, we expect the thin film to be in the mixed-state phase, with vortices present in the superconducting material. 
Ref.~\cite{TR_HINDS_VORTEXNOISE_07} showed that the random hopping of a vortex line from a pinning site to another could affect the trap lifetime. We stress here that, in addition, the motion of the vortex line submitted to a RF field is in itself responsible for dissipation, a phenomenon known as ``flux-flow''. However, this motion is partially suppressed because of the pinning of the vortices on material defects. Typical spin-flip frequencies are significantly smaller than vortex pinning characteristic energies (in the \giga\hertz\usk range). Many theoretical and experimental studies have already been carried out on the dissipation of type-II superconductors in this low-frequency regime. We note that observations are accounted for only if one assumes a non-local response of the material to an applied electromagnetic field. We present in this letter an adaptation of the theoretical framework developed in Ref.~\cite{TR_HINDSREKDAL_FLIPCALCULATION_04} to this particular non-local situation. We have adapted the theory describing the vortex-dynamics in niobium~\cite{LPA_PRLVORTEXDEPINNING_97, LPA_VORTEXDYNAMIC_98, LPA_EPLSLIPPAGELENGTH_04} to the situation of atom-chips. On the basis of the measurements of the previous references, we evaluate quantitatively the influence of vortex dissipation on atomic lifetime for the particular case of our superconducting atom-chip  experiment~\cite{ENS_CHIPSUPRA06,ENS_BECSUPRA08}. 

\section{Spin-flip lifetime calculation}

Let us first consider the simple case of an infinite thickness conducting slab ($h \rightarrow \infty$). The atom can decay towards an untrapped state $|f\rangle$, $\omega$ being the frequency of the $i\rightarrow f$ transition. 
We define $k_0=\omega / c$. The contribution of a semi-infinite space  to the spin-flip rate can be calculated in term of the field Green's functions~\cite{TR_HINDSREKDAL_FLIPCALCULATION_04} which is equivalent to evaluating the field radiated by the atom onto itself~\cite{ENS_HOUCHES90}. This field can be decomposed into propagating and evanescent plane waves (Weyl decomposition). Each of these waves is reflected by the surface according to Fresnel laws before going back to the atom. One obtains~\cite{TR_REKDALSUPERCONDNOISE06}:
\begin{eqnarray}\label{eq:Weyldecomp}
 & & \Gif^{\mbox{\tiny slab}}(\om)  =  \Gif^0(\om) (n_{th}+1)  \dfrac{3}{8} Re \left(  \int_0^\infty dq \dfrac{q}{\eta_0(q)} \right.\\ 
 & & \left. \times  e^{2i \eta_0(q) k_0 d} \left[ r_p(q) - \eta_0^2(q) r_s(q)+2 q^2 r_s(q) \right] \right), \nonumber
\end{eqnarray} 
where $\Gif^0(\om)= \mu_0 (\mu_B g_S)^2 k_0^3/(24 \pi \hbar)$ is the spin-flip rate in vacuum, $n_{th}=1/(e^{\frac{\hbar \om}{k_B T}}-1)$ is the mean photon number at frequency $\om$, $\mu_B$ is the Bohr magneton and $g_S$ the gyromagnetic factor of the electron. The integration factor $q$ is such that $q k_0$ is the modulus of the wave vector component parallel to the surface. Evanescent waves correspond to $q > 1$. 

If we now consider the case a material described by a local dielectric permittivity $\varepsilon(\omega)$, the polarization-dependent Fresnel coefficients are:
\begin{equation}\label{eq:Fresnelvseps}
 r_s(q) = \dfrac{\eta_0(q) - \eta(\om,q)}{\eta_0(q) + \eta(\om,q)}, \quad r_p(q) = \dfrac{\varepsilon(\om)\eta_0(q) - \eta(\om,q)}{\varepsilon(\om)\eta_0(q) + \eta(\om,q)}
\end{equation} 
where $\eta_0(q) = \sqrt{1-q^2}$ and $\eta(q,\om)=\sqrt{\varepsilon(\om)-q^2}$. For a metal described by the Drude model, the permittivity, much larger than 1,  is related to the local conductivity $\sigma$: $\varepsilon(\om) = 1 + i \sigma / (\varepsilon_0 \om) \approx i \sigma / (\varepsilon_0 \om)$. The characteristic length associated with the material response is the skin-depth $\delta=\sqrt{2/(\mu_0 \sigma \om)}$, typically in the \micro\metre~range for good conductors at rf frequencies. The semi-infinite slab assumption holds then for $h \gg \delta$. 

In the near field regime, we have $1/k_0 \gg d$. Moreover if we make the experimentally reasonable assumption $d \gg \delta$, the main contribution to the integral of Eq.~(\ref{eq:Weyldecomp}) is for $q$ values such that $1 \ll q \ll \sqrt{|\varepsilon(\om)|} $. One thus obtains an asymptotic expression of Eq.~(\ref{eq:Weyldecomp}) with $\eta_0 \approx i q$ and $\eta(\om) \approx \sqrt{\varepsilon(\om)}$ leading to an analytical expression for $\Gif$~\cite{TR_HINDS_LIFETIMECONSUPERCON_05}:
\begin{eqnarray}
  \Gif(\om) & \approx & \Gif^0 (n_{th}+1)\left( 1+ \dfrac{27}{64} \dfrac{Re \left[ \dfrac{2}{\sqrt{\varepsilon(\om)}} \right]}{k_0^4 d^4}   \right) \label{eq:lifetimevseps}\\
   & \approx & \Gif^0 (n_{th}+1)\left( 1+ \dfrac{27}{64} \dfrac{\delta}{k_0^3 d^4}\right)\label{eq:lifetimemetal},
\end{eqnarray} 

For the case of superconductors, different theoretical models for the material response have been used in order to evaluate the spin-flip rate~\cite{TR_REKDALSUPERCONDNOISE06, TR_HINDSELIASHBERG07, TR_REKDAL_ELIASHBERG_07}: the phenomenological two-fluid model, the BCS microscopic model, and the Eliashberg theory which takes into account the scattering of Cooper pairs by the phonons. All those models predict a local complex conductivity $\sigma = \sigma_1 + i\sigma_2$. As soon as the temperature $T$ is significantly below the critical temperature $T_c$, one has $\sigma_2 \gg \sigma_1$. Equation (\ref{eq:lifetimevseps}) then becomes~\cite{TR_REKDALSUPERCONDNOISE06}:
\begin{equation}\label{eq:lifetimesupercon}
 \Gif(\om) \approx \Gif^0(\om) (n_{th}+1)\left( 1+ \dfrac{27}{64} \dfrac{1}{\sqrt{\om \mu_0}k_0^3 d^4} \dfrac{\sigma_1}{\sigma_2^{3/2}}\right).
\end{equation} 
The spin-flip rate is then reduced by more than 6 orders of magnitude as compared to the case of normal metal at similar temperature.

\section{Adaptation to the case of a type-II superconductor}

We turn now to the situation where vortices are present in the film. In order to find a relation between the vortex dynamics and the electromagnetic radiation, we assume that the atom-surface distance is much larger than the intervortex distance $a_0$. In this situation, the response of the vortex lattice to an electromagnetic field can be treated like that of a continuous complex hydrodynamic system, taking into account vortex pinning as well as vortex interactions~\cite{MATHIEUSIMON_GLMACRO_88, SONIN_2MODES_92, PLACAIS_2MODES_96}. The theoretical models that consider a local response of the mixed-state to the electromagnetic field~\cite{MX_LOCALVORTEXCLEM_91,MX_LOCALVORTEXBRANDT_91} underestimate the vortex dissipation at low frequency. It is necessary to consider a non-local response of the superconductor~\cite{LPA_PRLVORTEXDEPINNING_97,LPA_VORTEXDYNAMIC_98} which makes it impossible to define a local dielectric constant or a local conductivity for describing the material. Hence Eqs.~(\ref{eq:Fresnelvseps}-\ref{eq:lifetimesupercon}) do not apply directly. 

In the case of a non-local response of the superconductor, dissipation is well described in term of surface impedance $Z_S=\mu_0E_S/B_S$ at frequency $\om$, where $E_S$ and $B_S$ are the tangential electric and magnetic fields   on the surface respectively. The use of $Z_S$ allows to include the detailed microscopic response of the superconducting medium into a linear local relation between the tangential electric field at the surface and the surface current $\ve{K} = \ve{E}_{S}/Z_S$. 
The rate $\Gif$ being related to the dissipation in the material,  we expect it to be proportional to $Re(Z_S)$. 

%
%

In order to find the relation between $\Gif$ and $Z_S$, we express the Fresnel coefficients in terms of surface impedance:
\begin{equation}\label{eq:FresnelvsZ}
 r_s(q) = \dfrac{\eta_0(q) Z_S - Z_0}{\eta_0(q) Z_S + Z_0}, \quad r_p(q) = \dfrac{Z_0 \eta_0(q) - Z_S}{Z_0 \eta_0(q) + Z_S}
\end{equation} 
where $Z_0 = \mu_0 c$ is the vacuum impedance. For the radiation of a dipole at a distance $d$ the wave vector amplitude  $k_0q$ values mainly contributing to the integral in Eq.~\ref{eq:Weyldecomp} are of the order of $d^{-1}$. For all reasonable superconductor models and for the range of distances considered above ($d \gg$\unit{1}{\micro\metre}), these wave vectors are much smaller than the inverse of the field penetration depth at the frequency $\om$. The surface impedance $Z_S$ is then independent of $q$ and equal to its value at normal incidence. Substituting Eq.~\ref{eq:FresnelvsZ} into Eq.~\ref{eq:Weyldecomp} and performing the integration is equivalent to the substitution $\sqrt{\varepsilon(\om)}=Z_0/Z_S$ in Eq.~\ref{eq:lifetimevseps}:
\begin{equation}\label{eq:lifetimeZs}
\Gif(\om) \approx \Gif^0(\om) (n_{th}+1)\left[ 1+ \dfrac{27}{64} \dfrac{2}{\om \mu_0 k_0^3 d^4} Re(Z_S)\right].
\end{equation}
Note that that if we replace the surface impedance by its standard value in the case of a normal metal $Z_S^{met} = (1-i)/(\sigma \delta)$, we exactly recover  Eq.~(\ref{eq:lifetimemetal}). 

Let us stress that, within our approximations, the surface impedance $Z_S$ allows to calculate exactly the field radiated by the atom in the $z>0$ half-space and hence the transition rate. Equation~(\ref{eq:lifetimeZs}) is thus valid even for a finite thickness slab. 

\section{Two-mode non-local response of the vortex lattice}

In order to derive $Z_S$ for a type-II superconducting material, we evaluate the vortex response to an external oscillating magnetic field. As we restrict ourselves to the situation where the distance $d$ is larger than the intervortex distance, it is possible to consider averaged macroscopic quantities for local electric and magnetic fields as well as for supercurrent densities. In this frame, the response of the superconductor can be derived from an equivalent of the Ginzburg-Landau free energy relating the macroscopic quantities~\cite{MATHIEUSIMON_GLMACRO_88}. For this purpose, it is necessary to introduce the vortex field $\ve{B_0}=n_V \varphi_0 \ve{\nu}$, where $\ve{\nu}$ is the local direction of the vortex lines, $n_V$ the density of vortices per unit area and $\varphi_0=h/2e$ is the quantum of flux. $\ve{B_0}$ is related to the macroscopic magnetic field $\ve{B}$ by the generalized London equation $\ve{B} + \mu_0 \lL^2 \ve{\nabla}\times\ve{V_s} = \ve{B_0}$, where $\ve{V_s}$ is the averaged macroscopic velocity of the Cooper-pairs and $\lL$ is the London length. In absence of vortices ($B_0=0$), one recovers the London equation which leads to the Meissner effect. In the mixed state, a fraction $\ve{B_0}$ of the applied magnetic field penetrates the film. In the presence of a non-uniform supercurrent ($\ve{\nabla} \times V_s \neq 0$) vortex and magnetic field lines do not coincide. 

\begin{figure}[htb]
 \centering
\includegraphics[width=8.5cm]{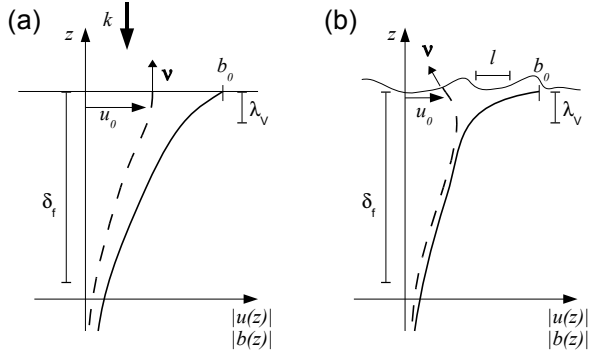}
  \caption{Amplitude of the transverse magnetic field $b$ (solid line) and displacement of a vortex line $u$ (dashed line): (a) in the case of a flat surface, (b) in the case of a rough surface with characteristic length $l$. For both situations, the vortex direction $\nu$ must end perpendicularly to the surface.}
 \label{fig:vortexsurf}
\end{figure}

We consider now the case of an electromagnetic field arriving at normal incidence on the superconducting medium according to the geometry of Fig.~(\ref{fig:vortexsurf}). The vortex lattice is initially in its equilibrium position $\ve{B_0}=\ve{B}=B \ve{e_z}$. The magnetic and vortex fields experience a small perturbation $\ve{B}(z)= B \ve{e_z} + b_0 exp(i k z - i \om t) \ve{e_x}$, $\ve{\nu}(z)= \ve{e_z} + \nu_0 exp(i k z - i \om t) \ve{e_x}$, where $|\nu_0|$ is the angle of the vortex line with the $z$-direction on the surface and $|\nu(z)|$ the same angle at the position $z$.  Refs.~\cite{SONIN_2MODES_92,PLACAIS_2MODES_96} show that the field in the medium is a superposition of two propagation modes with wave vector $k_f$ and $k_V$. The characteristic length of penetration $\delta_f$ associated to the first mode is related to the flux-flow dissipation when a vortex line is moving:
\begin{equation}\label{eq:lambdaf}
 \delta_f = \sqrt{\dfrac{2}{\mu_0 \sigma_f \om}} \approx \sqrt{\dfrac{2 \rho_N}{\mu_0 \om}\dfrac{B}{B_{c_2}}}.
\end{equation}  
$B_{c_2}$ is the second critical magnetic field of the material, and $\rho_N$ is the resistivity of normal electrons in  the material at temperature $T$. As a consequence $\delta_f$ is of the order of typical skin-depths for metals at low temperature. We define the complex penetration length $\lambda_f=1/i k_f =\delta_f(1+i)/2$.

The characteristic penetration length for the second mode is:
\begin{equation}\label{eq:lambdaV}
 \lambda_V = \dfrac{1}{ik_V} = \lL \sqrt{\dfrac{\mu_0 \epsilon_l}{B+\mu_0 \epsilon_l}}.
\end{equation}  
where $\epsilon_l$ is the vortex line potential~\cite{MATHIEUSIMON_GLMACRO_88}. The quantity $\varphi_0 \epsilon_l$ is the energy that is required to enter one unit of length of vortex into the medium. The penetration length $\lambda_V$ is of the order of $\lambda_L$. In usual superconductors one has $\lambda_V \ll \delta_f$. The second mode describes the non dissipative-screening of the magnetic field by supercurrents located close to the surface.

The incoming field must be decomposed onto these two modes of propagation. For each mode we can define the displacement of the vortex line at the surface $u_{j=(f,V)}=\int_{-\infty}^0 \nu_j(z) dz$. The relative weight of each mode is determined by the boundary conditions at the surface for the magnetic and vortex fields. It strongly depends on the surface geometry of the sample. We present in Fig.~\ref{fig:vortexsurf} two different cases of propagation in the superconducting medium for a perfectly flat (a) and a rough surface (b). In both cases the vortex field must end perpendicularly to the surface. In the second case its displacement is  prevented by its pinning on surface defects. It was proposed in Refs.~\cite{LPA_PRLVORTEXDEPINNING_97, SONIN_2MODES_92} to link the vortex displacement at the surface $u_0 = u_f+u_V$ to the angle $\nu_0$ by the relation $u_0 + l \nu_0 = 0$, where $l$ is the phenomenological slippage length that characterizes the material. Following Fig~\ref{fig:vortexsurf}(b), one clearly sees that $l$ is related to the roughness of the surface but it also takes into account the interaction between vortices that forces a collective response of the whole lattice~\cite{LPA_EPLSLIPPAGELENGTH_04}. 

The complete determination of the surface magnetic and electric fields, together with the vortex displacement is presented in Refs.~\cite{SONIN_2MODES_92,PLACAIS_2MODES_96} for a infinite half-space. The resulting surface impedance is:
 \begin{equation}\label{eq:ZSinfinity}
  Z_S^\infty  =  -i \mu_0 \om \dfrac{B (l +\lambda_V) \lambda_f}{(B+\mu_0 \epsilon_l)(l+\lambda_V)+\mu_0 \epsilon_l \lambda_f}
 \end{equation}

\section{Finite-thickness effects}

As shown below, in our experimental conditions, the slab thickness $h=$\unit{1}{\micro\metre} is of the order of or smaller than $\delta_f$. The superconducting medium thus cannot be described as a half-space and finite size effects have to be taken into account.
In the case of a type-II superconductor, they have been studied both theoretically and experimentally~\cite{LPA_VORTEXDYNAMIC_98}, in a regime where the magnetic field is the same on each side of the slab. It does not correspond to our experimental situation because the presence of the magnetic dipole breaks the symmetry between the two sides of the slab.

\begin{figure}
\begin{center}
 \includegraphics[width=8cm]{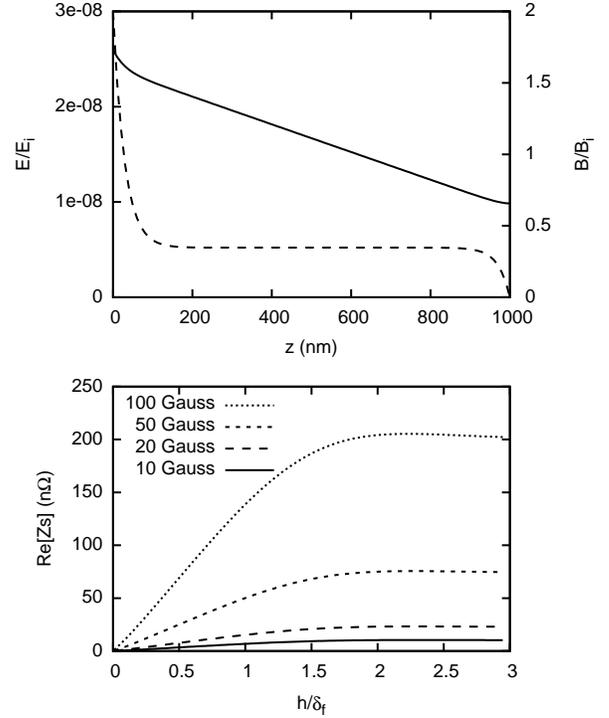}
\end{center}
\caption{(a) Electric (solid line) and magnetic (dashed line) field amplitudes in the superconducting slab of thickness $h=$\unit{1}{\micro\metre}. The superconductor characteristic parameters $\delta_f$, $\lambda_V$ and $l$ correspond to the niobium layer used in our experiment (see section ``Numerical results''). The bias field applied perpendicularly to the slab is $B=100$~G. The fields are expressed in units of the incident electromagnetic field amplitude. (b) Real part of the surface impedance as a function of $h$ (in units of $\delta_f$) for different external bias field $B$}
\label{fig:finitethickness}
\end{figure}

We have calculated the surface impedance $Z_S$ in the case of the reflection of an incident electromagnetic wave arriving at normal incidence on a superconducting slab of finite thickness. It requires to take into account 7 modes of propagation: the incident, reflected and transmitted fields as well as the evanescent propagation modes in the film with wave vectors $\pm k_f$ and $\pm k_V$. Figure \ref{fig:finitethickness}(a) presents the analytical results for the variation of the electric and magnetic fields inside the superconducting film. We observe that only a very small part, of the order of $10^{-8}$, of the incident field penetrates the slab, as expected for any metal. Moreover, most of the magnetic field is screened  by superconducting currents carried by the the modes $\pm k_V$ which do not dissipate. Hence an even smaller fraction of the incoming wave is dissipated by the vortex displacement and contributes to $Re(Z_S)$. 

We present in Fig.~\ref{fig:finitethickness}(b) the calculated real part of $Z_S$ as a function of the slab thickness for different external field values $B$. We recover the value $Re(Z_S^\infty)$ of Eq.~(\ref{eq:ZSinfinity}) as soon as $h\gtrsim 2\delta_f$. In our experiment, we are in the opposite limit $h\ll \delta_f$, for which we fit the results with the phenomenological formula:
\begin{equation}\label{eq:Zfinite}
Re[Z_S(h)] \approx \dfrac{2}{3}\dfrac{h}{\delta_f}Re(Z_S^\infty) 
\end{equation}

%
%

\section{Numerical results}
\label{sec:numericalresults}
In the specific case of our superconducting atom-chip~\cite{ENS_CHIPSUPRA06,ENS_BECSUPRA08}, we can evaluate all relevant parameters and determine $Z_S$. Resistance measurements of the sputtered Nb film give a normal resistivity above transition to the superconducting state $\rho_N=$\unit{15}{\micro\ohm\usk\centi\metre}. The comparison with measurements at room temperature gives a residual resistance ratio (RRR) of 4.6, which indicates that the film is in the so-called ``dirty limit''. Measurements of the critical magnetic field for similar films in this limit gives $B_{c_2}=4.5\cdot 10^4$~G, a factor 15 larger than the pure case value~\cite{PEROZ_NBCLEANDIRTY_05}. To first order the product $B_{c_1}B_{c_2}$ is a quantity that weakly depends on the quality of the film and remains almost constant~\cite{TX_DEGENNESSUPERCONDUCTIVITY}. Hence, we expect $B_{c_1}$ to be a factor 15 smaller than the pure case value~\cite{FINNEMORE_NBPURBC1_66}, i.e. $B_{c_1}= 80$~G. The vortex potential $\epsilon_l$ depends on the applied magnetic field $B$. For normal operation of an atom-chip, we fulfill the condition $B\approx B_{c_1} \ll B_{c_2}$ and we have $\mu_0 \epsilon_l \approx 0.9 B_{c_1} \approx 70$~G~\cite{MX_ABRIKOSOV57}. 

A microscopic calculation of $l$ is out of the scope of this Letter. It can be found in~\cite{LPA_EPLSLIPPAGELENGTH_04}.  Nevertheless, it is possible to evaluate this quantity by linking it to the critical current in the superconducting slab $I_c=\iint j_s d^2r$. On the one hand, we have $I_c \approx 2 w \epsilon_l \nu_c$~\cite{LPA_VORTEXDYNAMIC_98}, where $w \gg h$ is the width of the slab and $\nu_c$ is the maximum angle at which the vortex line can bend before their pinning on the surface breaks. On the other hand, the critical angle is reached when the displacement $u$ is of the order of the intervortex separation $a_0 = \sqrt{\varphi_0 / B}$. Combining those two equations with the boundary condition between $u_0$ and $\nu_0$, we obtain:
\begin{equation}\label{eq:lvsIc}
 l = \dfrac{\epsilon_l w}{I_c}\sqrt{\dfrac{\varphi_0}{B}}.
\end{equation} 

\begin{figure}

 \centering
 \includegraphics[width=8cm]{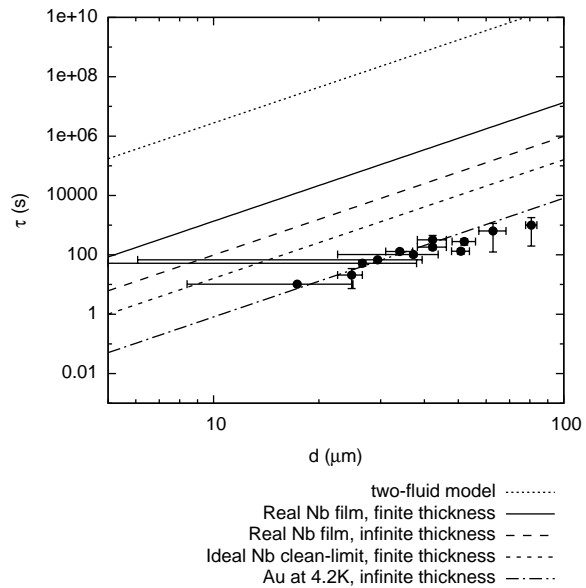}
 \caption{Spin-flip lifetime as a function of atom-surface distance as predicted by Eq.~(\ref{eq:lifetimeZs}) (solid line). It takes into account the measurements of our real superconducting film properties and its finite thickness $h=$\unit{1}{\micro\metre}. The dashed lines correspond to two ideal cases where the thickness of the film is infinite (long dashes), or its superconducting properties are ideal (clean limit, short dashes). The dotted line corresponds to the superconducting BCS model~\cite{TR_HINDSELIASHBERG07}. The experimental points correspond to our measurements for a Gold layer of thickness $h=200$~\nano\metre~\cite{ENS_EXPLIFETIMEGOLD08}. They are in good agreement with the predictions of Eq.~(\ref{eq:lifetimemetal}), with $\sigma_{Au,4.2K}=6.7\times 10^9$~\ohm\usk\metre (dot-dashed line)}
 \label{fig:tau_vs_d}
\end{figure}

We will now consider that the atoms are trapped above the $h=$\unit{1}{\micro\metre} thick superconducting Z-wire of Ref.~\cite{ENS_CHIPSUPRA06}. We consider this finite width conductor ($w=$\unit{40}{\micro\metre}) as an infinite film, leading to a worst case estimate for the spin-flip rate. We assume that an homogeneous external bias field $B=100$~G is applied perpendicularly to the slab. This crude assumption corresponds to a worst case situation. Another external magnetic field parallel to the surface determines the spin-flip frequency $\om=2 \pi \times 2$~\mega\hertz. Eq.~(\ref{eq:lambdaf}) gives then $\delta_f =$\unit{9.2}{\micro\metre} and using Eq.~(\ref{eq:lambdaV}) we obtain $\lambda_V = 29$~\nano\metre~(with $\lambda_L =45$~\nano\metre~\cite{FINNEMORE_NBPURBC1_66}). In presence of a current in the slab, the value $l$ can be obtained by replacing $I_c$ by $I_c-I$ in Eq.~(\ref{eq:lvsIc}) where $I=1.4$~\ampere~is the actual current in the slab required for trapping ($I_c=1.76$~\ampere). We get $l=250$~\nano\metre. Figure~\ref{fig:tau_vs_d} compares the spin-flip lifetime $\tau=\Gif^{-1}$ derived from Eqs.~(\ref{eq:Zfinite}) and (\ref{eq:lifetimeZs}) as a function of the distance $d$. These predictions are compared to our measurements for a gold layer of thickness $h=200$~\nano\metre~\cite{ENS_EXPLIFETIMEGOLD08}. We also present the results of the BCS model of Refs.~\cite{TR_REKDALSUPERCONDNOISE06,TR_HINDSELIASHBERG07,TR_REKDAL_ELIASHBERG_07}. 

Our predictions stand in a regime where the distance $d \gg \delta_f, \lambda_V, a_0$. The results of Fig.~\ref{fig:tau_vs_d} are therefore valid for $d\gtrsim$\unit{10}{\micro\metre}. In this regime the vortex dissipation is the limiting factor for the spin-flip lifetime. We find a reduction of about 3 orders of magnitude of the lifetime with respect to the two-fluid model. However, superconducting Nb remains significantly better than normal metals. 

We have also represented in Fig.~\ref{fig:tau_vs_d} the predicted lifetimes in the case of an semi-infinite superconducting film, where the surface impedance is given by Eq.~(\ref{eq:ZSinfinity}), or for a film with ideal purity and surface quality (clean limit). It is interesting to note that these two situations correspond to a degraded lifetime compared to the dirty film. This might seem paradoxical. In this clean superconductor regime however, surface pinning is significantly reduced, the vortices can move with a larger amplitude. In addition the change of the normal fluid conductivity increases the viscous drag of the vortex lattice. Two phenomenons therefore contribute to the increase of the dissipation. 

We also note that we have considered here a worst case limit where we assume that all the surface is subjected to a an external field of 100~G and a homogeneously distributed current 1.4~\ampere. A more precise calculation of the the atomic losses should include the inhomogeneous current and vortices distribution on the surface. Nevertheless, our results show that vortices play a crucial role in the spin-flip lifetime of atoms in the close vicinity of a type-II atom-chip.

In conclusion we have adapted the formalism of atomic spin-flip lifetime of an atom close to a metallic surface to the non-local electrodynamic response of the vortex lattice in type-II superconductors. Note that this model should also described type-I superconductors, as a thin film of such a material will contain vortices. On the other hand, lifetime close to superconducting materials remains significantly better than close to normal metals. Our results predict a lifetime of \unit{10000}{\second} at \unit{20}{\micro\metre} from the surface opening new perspectives for the coherent manipulation of ultracold atoms in the vicinity of superconductors. 

\acknowledgments
Laboratoire Kastler Brossel and Laboratoire Pierre Aigrain are joint research Laboratories of CNRS with \'Ecole normale sup\'erieure and Universit\'e Pierre et Marie Curie. EM2C is a laboratory of CNRS associated to \'Ecole centrale de Paris. We acknowledge support of the European Union (CONQUEST and SCALA projects, Marie Curie Fellowship program), of the Japan Science and Technology corporation (International Cooperative Research Project~: ``Quantum Entanglement''), of the ANR, DGA and of the R\'egion Ile de France (IFRAF and Cnano Idf consortiums). 

\bibliographystyle{eplbib}

\end{document}